\documentclass[%
 aip,
 sd,%
 amsmath,amssymb,
 reprint,%
]{revtex4-1}

\usepackage{graphicx}
\usepackage{dcolumn}
\usepackage{bm}

\begin{document}

\title{Exciton valley dynamics in monolayer WSe$_2$ probed by the two-color ultrafast Kerr rotation}

\author{Tengfei Yan, Jialiang Ye, Xiaofen Qiao, Pingheng Tan, and Xinhui Zhang}
 \altaffiliation[]{ ~State Key Laboratory of Superlattices and Microstructures, Institute of Semiconductors, Chinese Academy of Sciences, P.O. Box 912, Beijing 100083, People's Republic of China. E-mail: xinhuiz@semi.ac.cn.}

\begin{abstract}
The newly developed two-dimensional layered materials provide perfect platform for valley-spintronics exploration. To determine the prospect of utilizing the valley degree of freedom, it is of great importance to directly detect and understand the valley dynamics in these materials. Here, the exciton valley dynamics in monolayer WSe$_2$ is investigated by the two-color pump-probe magneto-optical Kerr technique. By tuning the probe photon energy in resonance with the free excitons and trions, the valley relaxation time of different excitonic states in monolayer WSe$_2$ is determined. Valley relaxation time of the free exciton in monolayer WSe$_2$ is confirmed to be several picoseconds. A slow valley polarization relaxation process is observed to be associated with the trions, showing that the valley lifetime for trions is one order of magnitude longer than that of free excitons. This finding suggests that trion can be a good candidate for valleytronics application.
\end{abstract}

\maketitle

Monolayers of transition metal dichalcogenides (TMDCs) such as MoS$_2$ and WSe$_2$ are newly developed two-dimensional materials with the direct bandgap in the visible region at the energetically degenerate $K^+$ and $K^-$ points of the hexagonal Brillouin zone\cite{AtomicallythinMoS2anewdirectgapsemiconductor,EmergingphotoluminescenceinmonolayerMoS2,Progresschallengesandopportunitiesintwodimensionalmaterialsbeyondgraphene,DirectobservationofthetransitionfromindirecttodirectbandgapinatomicallythinepitaxialMoSe2,Giantspinorbitinducedspinsplittingintwodimensionaltransitionmetaldichalcogenidesemiconductors}. The optical properties of monolayer TMDCs are strongly influenced by the excitonic effects because there exists strong Coulomb interaction between electrons and holes due to the strict two-dimensional limit and the screening effect. Optical response is dominated by both exciton, \emph{i.e.}, the combination of a pair of confined electron and hole, and trion, in another name, charged exciton, even at room temperature\cite{Electricalcontrolofneutralandchargedexcitonsinamonolayersemiconductor,Groundstateenergyandopticalabsorptionofexcitonictrionsintwodimensionalsemiconductors,Largeexcitoniceffectsinmonolayersofmolybdenumandtungstendichalcogenides,OpticalgenerationofexcitonicvalleycoherenceinmonolayerWSe2,Opticalmanipulationoftheexcitonchargestateinsinglelayertungstendisulfide,TightlyboundtrionsinmonolayerMoS2,OpticalSpectrumofMoS2ManyBodyEffectsandDiversityofExcitonStates}. Moreover, the evidence of the existence of biexciton in monolayer TMDCs has also been recently reported\cite{you2015observation}. The abundant excitonic effects indicate the potential applications in novel photonic and optoelectronic devices based on the atomically thin TMDC materials.

\begin{figure}[!htb]
	\includegraphics[width=8.5cm]{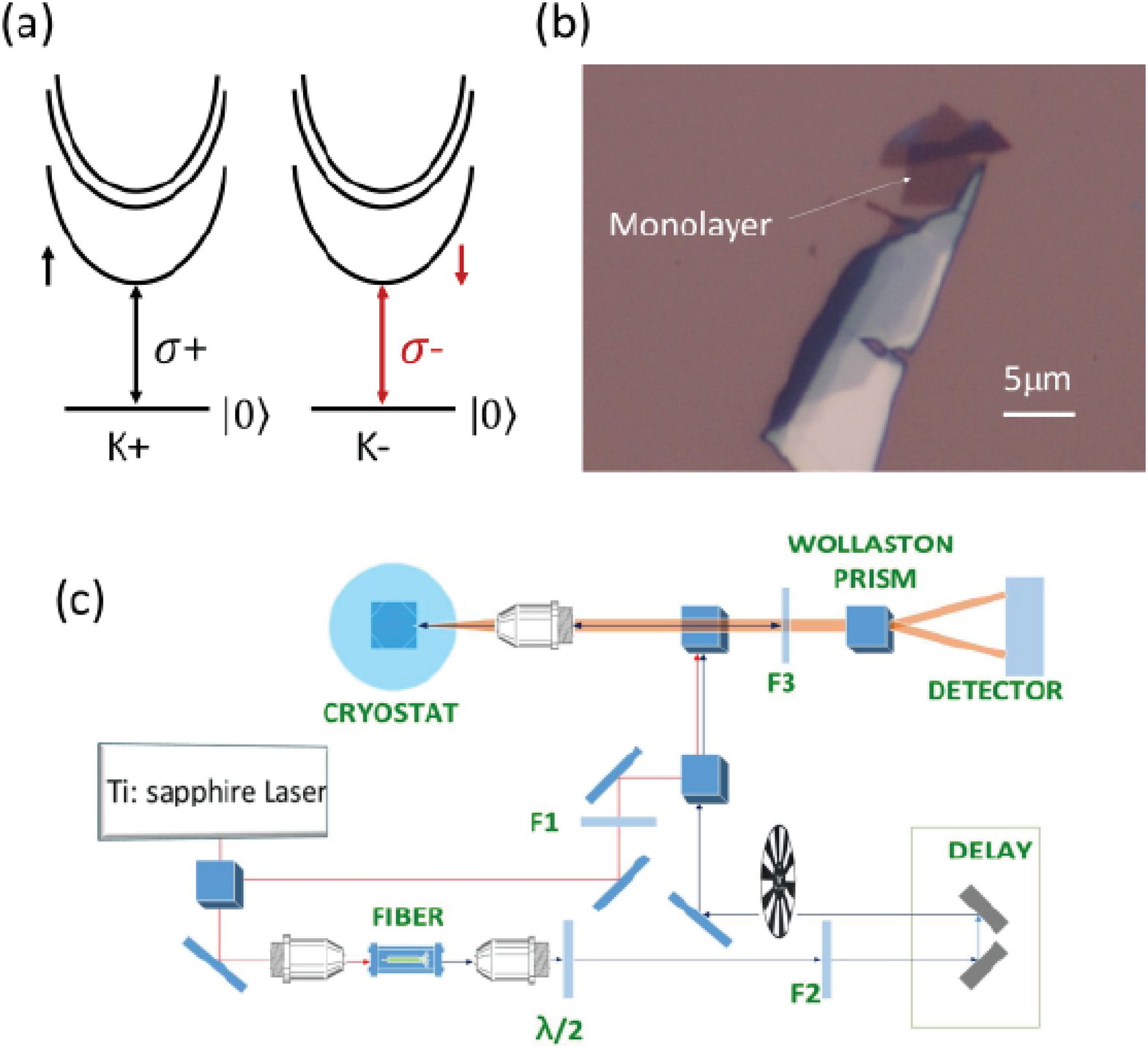}
	\caption{(a) Optical valley selection rule for the exciton in monolayer WSe$_2$. (b) Optical microscopic image of pre-marked WSe$_2$ monolayer flake. (c) Sketch of the time-resolved Kerr rotation measurement setup, with F1, F2 and F3 representing optical filters respectively.}
	\label{1}
\end{figure}

The bright excitons and trions in monolayer TMDCs are located at the unequivalent $K^+$ or $K^-$ valley, thus possess additional degree of freedom (DoF), valley DoF. Thanks to the broken inversion symmetry and large spin-orbit coupling, monolayer TMDCs characterize an optical selection rule, which describes the lock between the light circular dichroism and the exciton valley as well as spin DoF\cite{Valleydependentoptoelectronicsfrominversionsymmetrybreaking,CoupledspinandvalleyphysicsinmonolayersofMoS2andothergroupVIdichalcogenides}, as shown in FIG. \ref{1}a. Previous works have realized the exciton valley polarization injection and detection using the steady-state helicity-resolved photoluminescence (PL) technique \cite{Valleyselectivecirculardichroismofmonolayermolybdenumdisulphide,ValleypolarizationinMoS2monolayersbyopticalpumping,ControlofvalleypolarizationinmonolayerMoS2byopticalhelicity,ValleypolarizationandintervalleyscatteringinmonolayerMoS2,RobustopticalemissionpolarizationinMoSmonolayersthroughselectivevalleyexcitation,ValleydepolarizationinmonolayerWSe2}. The exciton valley dynamics was then studied by the helicity- and time-resolved PL\cite{CarrierandPolarizationDynamicsinMonolayerMoS2,ValleydynamicsprobedthroughchargedandneutralexcitonemissioninmonolayerWSe2,yang2015long} and reflectance spectrum\cite{doi:10.1021/nl403742j,doi:10.1021/nn405419h,C4NR03607G,ExcitonvalleyrelaxationinasinglelayerofWS2measuredbyultrafastspectroscopy}. Recently, the time-resolved Kerr rotation (TRKR), a technique which directly probes the valley or spin polarization in the material, has been applied to investigate the exciton valley dynamics in TMDCs\cite{PhysRevB.90.161302,plechinger2014time}. Exciton valley relaxation time was confirmed to be in the picosecond range. But more detailed studies, especially the trion valley lifetime and its correlation with that of excitons, are still needed, in order to uncover the valley relaxation process and push ahead with the development of valleytronics.

Here we use the two-color TRKR technique to investigate the valley dynamics in monolayer WSe$_2$. With the optical selection rule in monolayer TMDCs \cite{Valleycontrastingphysicsingraphenemagneticmomentandtopologicaltransport} as shown in FIG. \ref{1}a, the polarization of exciton spin state, which is correlated with a certain valley state, can be injected by the circularly polarized light excitation. The probe beam, which is linearly polarized, can be resolved into two circularly polarized light waves with equal amplitude and phase. The two circularly polarized light waves propagate with different refractive index in the sample with spin polarization, the Kerr rotation is then detected. Thus, we can detect the valley polarization in regardless of the quantity of the nonequilibrium carriers. By tuning the probe photon energy in resonance with free exciton and trion respectively, the valley relaxation times of excitonic states in picosecond range at low temperatures are deduced. A relatively slow relaxation process of the residual valley polarization for trion is observed after the relaxation of neutral exciton states, suggesting that trion is a good candidate for valleytronics application.

\section{Experimental Section}

The WSe$_2$ flakes are fabricated by the mechanical exfoliation with adhesive tape from a bulk crystal (2D semiconductors Inc.) onto $SiO_2/Si$ substrates. Few-layer and monolayer WSe$_2$ flakes were first identified by optical contrast under a microscope (see FIG. \ref{1}b), then verified via Raman\cite{:/content/aip/journal/apl/106/22/10.1063/1.4921911} and PL measurements, using the method presented in our previous work\cite{:/content/aip/journal/apl/105/10/10.1063/1.4895471}. The steady-state PL is collected by a home-made micro-PL system based on a Horiba Jobin Yvon iHR550 spectrometer equipped with a Synapse Si CCD camera
. 
For the circular polarization resolved experiment, the PL emission passes through a quarter wave plate and a beam-displacer for the \emph{s}- and \emph{p}-polarized components detection at the same time.

The TRKR experiments are performed with the optical setup shown in FIG. \ref{1}c. The pump beam is fixed at 1.78 eV from a femtosecond pulsed laser (Chameleon, Coherent, Inc.) and the probe beam is chosen from a supercontinuum white light using two bandpass filters (TBP01-790 and FF01-747, Semrock, Inc.) with the band width less than $\sim$10 meV. The super continuum white light is obtained from a nonlinear photonic crystal fiber (femotoWHITE-800, NKT Photonics) excited by the Ti:sapphire laser with a temporal pulse width being approximately 150 fs. The pump and probe beams are focused onto the sample collinearly using an Olympus 50$\times$ objective. The spatial resolution is about 1.5 $\mu$m. The intensity of the pump and probe beam is kept to be 100 $\mu$W and 20 $\mu$W during the experiment, respectively. The pump beam is circularly polarized and is cut off by a long-wavelength pass filter (FF01-692/LP, Semrock, Inc.) before being directed onto the photodetector. The probe beam is linearly polarized. A Wollaston prism is used to separate the reflected probe beam into two orthogonally polarized components, which are detected by a silicon balanced photodetector to achieve the weak Kerr rotation detection. All measurements are taken at low temperatures with the sample mounted in a microscopy cryostat (Janis Research Company).

\begin{figure}[!htb]
	\includegraphics[width=8cm]{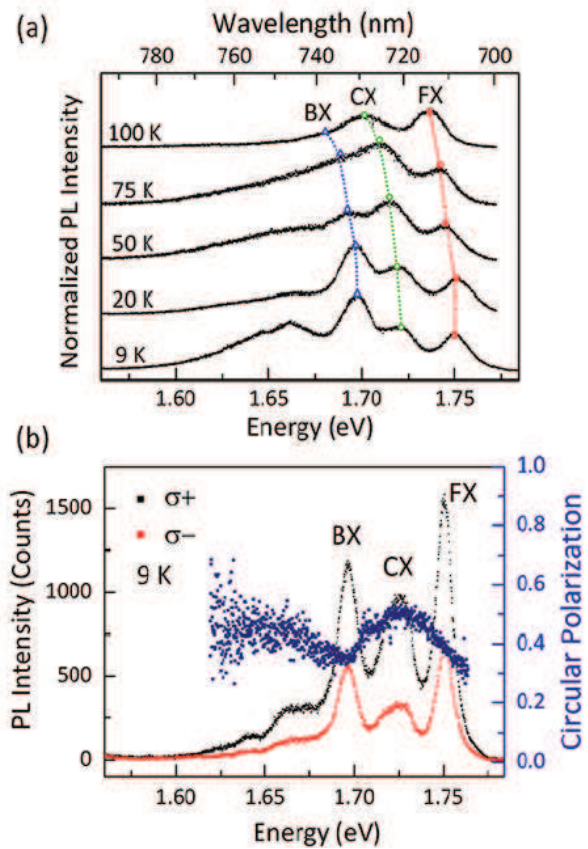}
	\caption{(a) Normalized PL response of monolayer WSe$_2$ at different temperatures. The dash lines show the temperature dependent energy shifts of the exciton PL peaks. (b) PL response and circular polarization degree of monolayer WSe$_2$ excited with pump photon energy of 1.78 eV at 9 K. The circular polarization degree is plotted in blue dots. }
	\label{2}
\end{figure}

\section{Results and Discussions}
The PL spectra of monolayer WSe$_2$ at different temperatures are shown in FIG. \ref{2}a, in which the excitonic emission peaks are clearly resolved. The emission peak labeled as FX and CX are related to the free exciton and trion recombinations, respectively. The peak labelled as BX may be related to the biexciton recombination, which locates about 53 meV lower than the free exciton emitting energy at 9 K, in consistent with the previous report\cite{you2015observation}. Emission peaks at even lower energy originate from the bandtail state and are beyond the study scope here\cite{:/content/aip/journal/apl/105/10/10.1063/1.4895471}. Whether the trions possess positive or negative excess charge is still yet to be determined. The trion bingding energy is deduced to be 29 meV from the PL spectrum, which is in consistent with the previous work\cite{OpticalgenerationofexcitonicvalleycoherenceinmonolayerWSe2,ValleydynamicsprobedthroughchargedandneutralexcitonemissioninmonolayerWSe2}, so that the trions can be well resolved at low temperatures and measurable even at room temperature. All the excitonic emission peaks show redshift with increasing temperature, and the thermal quenching of the trion and biexciton emission is also observed. The well-resolved trion state in monolayer WSe$_2$ at low temperature allows us to study its valley dynamics individually, which is quite different from the GaAs case whose binging energy is too small\cite{PhysRevB.79.233306,PSSA:PSSA489} to be resolved using a femtosecond pulsed laser.

The helicity-resolved PL measurement is investigated at 9 K. The monolayer WSe$_2$ flake is excited at normal incidence by a 1.78 eV laser beam with $\sigma^+$ helicity, while the PL with different circular helicities are collected separately\cite{ValleydepolarizationinmonolayerWSe2}, as shown in FIG. \ref{2}b. (There's some difference between the lineshapes in FIG. \ref{2}a and FIG. \ref{2}b at the same temperature, because of the increased excitation intensity used in the temperature dependent PL measurement.) The polarization degree of PL is calculated by the equation $P_c=(I_+-I_-)/(I_++I_-)$, where $I_+$ and $I_-$ correspond to the PL intensity of $\sigma^+$ and $\sigma^-$ components, respectively. The polarization degree for the neutral excitons including the free exciton and biexciton emission are evaluated to be nearly 40 percent, while the trion state possesses even higher valley polarization up to 50 percent. The steady-state PL results suggest that the trion state preserves the valley polarization more efficiently, thus being a more promising candidate for valley Hall experiment and future valleytronic applications \cite{ThevalleyHalleffectinMoStransistors}.

\begin{figure}[!htb]
	\includegraphics[width=7.5cm]{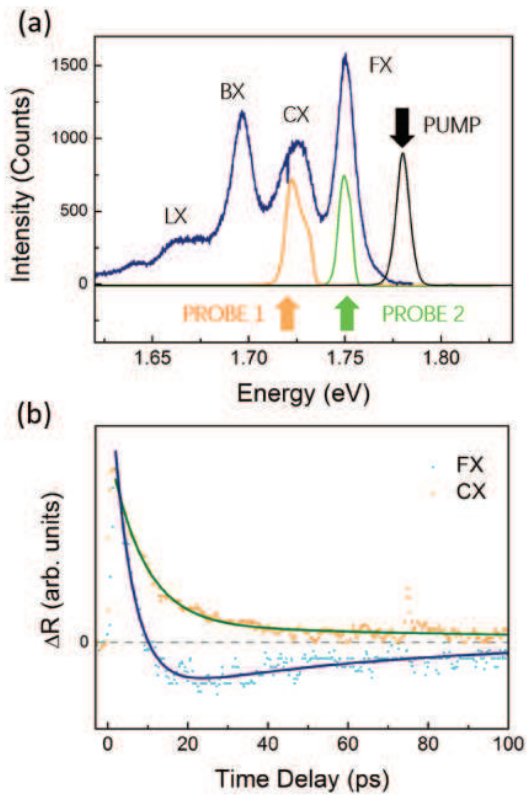}
	\caption{(a) The spectra of the selected pump and probe laser beams, with the photon energy of probe beam tuned to be resonant with the exciton and trion PL peaks (at 10 K), respectively, as indicated in the PL spectrum (bule curve). (b) The transient reflectance spectrum at 10 K measured at different probe photon energy, in resonance with FX (1.75 eV) and CX (1.72 eV), respectively. The solid lines are the exponential decay fitting results.}
	\label{3}
\end{figure}

The transient differential reflectance ($\Delta R$) response of the free exciton and trion state is then measured by tuning the probe beam energy to be 1.75 eV and 1.72 eV at 10 K, being in resonance with the free exciton and trion emission energy, respectively, as shown in FIG. \ref{3}a. The results indicate that both the exciton and trion decay in picosecond range, and could be fitted by bi-exponential decay functions.
The decay constants of the free exciton are deduced to be about 7$\pm1$ ps and 15$\pm4$ ps, while the trion decays with two time constants of 8$\pm0.5$ ps and 85$\pm8$ ps. It has to be noted that the deduced values of the decay constants depends on the experimental conditions such as the excitation power and photon energy which is beyond the scope of this work. The observed decay process of excitons is the complex consequence of exciton recombination and the rapid exciton-exciton annihilation according to the previous works\cite{sun2014observation,yuan2015exciton,pollmann2015resonant}.

\begin{figure}
	\includegraphics[width=7.8cm]{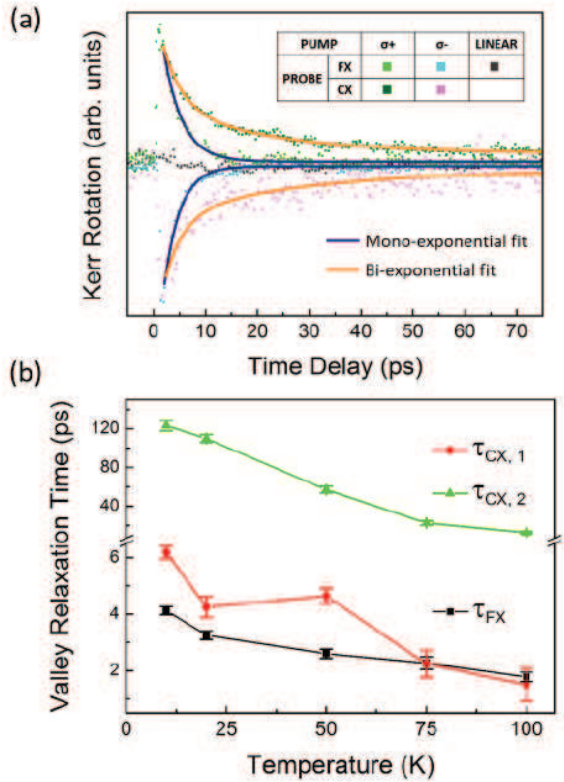}
	\caption{(a) The time-resolved Kerr rotation response measured at 10 K, with Kerr response pumped by $\sigma^+$ and $\sigma^-$, respectively, while the grey dots is the Kerr rotation when the sample is pumped by linearly polarized laser. The responses of free exciton and trion are shown in different colors. (b) The temperature dependent valley relaxation time, deduced from the time-resolved Kerr rotation traces, measured with the probe beam energy tuned to be in resonance with the FX or CX emitting peak according to the PL spectra at different temperatures.}
	\label{4}
\end{figure}

To further investigate the valley decay processes of different excitonic states, the TRKR measurement is performed and the results are shown in FIG. \ref{4}. The pump beam is fixed to be 1.78 eV and the probe beam energy is tuned to be in resonance with the free exciton and trion referring to the PL peak energy, respectively, same as the $\Delta$R measurement. The Kerr rotation signal flips its sign as the pump beam helicity changes from $\sigma^+$ to $\sigma^-$, and no Kerr response can be distinguished when pumped by the linearly polarized light, as can be seen in FIG. \ref{4}a. This indicates that all the signals originate from the imbalanced exciton valley occupation.

A monoexponential decay function is applied to fit the free exciton valley depolarization traces, and the decay time constant is deduced to be 4.1$\pm$0.2 ps at 10 K. This is in the same order of magnitude with the previously reported results\cite{PhysRevB.90.161302,plechinger2014time}. The free A exciton valley depolarization time of $\sim$6 ps in monolayer WSe$_2$ measured in a degenerate pump-probe geometry has been reported by Zhu \emph{et al.}\cite{PhysRevB.90.161302}, and in MoS$_2$, valley relaxation time is 17 ps as reported by Plechinger \emph{et al.}\cite{plechinger2014time}. Our two-color TRKR results indicate that the exciton valley lifetime is shorter compared to the exciton lifetime obtained by the transient reflectance spectrum. It is noted that the measured valley lifetime by TRKR is shorter than that evaluated by the PL polarization degree results under cw laser excitation. This results from the strong dependence of valley lifetime on the optically excited exciton density under the femtosecond laser excitation, as discussed in our previous study\cite{ValleydepolarizationinmonolayerWSe2}.

A systematic temperature dependent TRKR is performed. The probe photon energy is tuned to be in resonance with the temperature dependent (10$\sim$100 K) free exciton and trion emission energies, based on the results shown in FIG. \ref{2}a. It is seen that a general trend shows that the valley relaxation time decreases with increasing temperature, as shown in FIG. \ref{4}b, indicating an enhanced valley relaxation process at higher temperatures. This result is coincident with our previous PL results\cite{ValleydepolarizationinmonolayerWSe2}, in which the polarization degree of PL decreases with increasing temperature. The Maialle-Silva-Sham (MSS) mechanism caused by the electron-hole (\emph{e-h}) exchange interaction\cite{Excitonspindynamicsinquantumwells,OpticalgenerationofexcitonicvalleycoherenceinmonolayerWSe2,PhysRevB.90.161302,ValleydepolarizationduetointervalleyandintravalleyelectronholeexchangeinteractionsinmonolayerMoS,Excitonfinestructureandspindecoherenceinmonolayersoftransitionmetaldichalcogenides,ValleydepolarizationinmonolayerWSe2} has been proposed to be responsible for the rapid exciton valley relaxation in monolayer TMDCs as theoretically pointed out in the previous studies\cite{ValleydepolarizationduetointervalleyandintravalleyelectronholeexchangeinteractionsinmonolayerMoS,Excitonfinestructureandspindecoherenceinmonolayersoftransitionmetaldichalcogenides}. The long-range \emph{e-h} exchange interaction acts as a momentum-dependent effective magnetic field, in analogy to the spin-orbit coupling in GaAs spin relaxation\cite{Spindynamicsinsemiconductors}. The valley pseudospin of excitons with different center-of-mass momentums precess around the effective magnetic field with different frequencies, leading to a free-induction decay as that of polarization relaxation induced by the inhomogeneous broadening caused by the randomly-orientated effective magnetic field. It is thus expected to observe a shorter valley lifetime at higher temperatures, because of the enhanced \emph{e-h} exchange interaction.

The valley relaxation for trions is more complicated, two decay time constants are deduced by fitting the TRKR valley depolarization trace with a biexponential decaying function, being 6$\pm$0.2 ps and 123$\pm$5 ps at 10 K, respectively. The fast decay constant is in the similar magnitude with that of exciton in the whole measurement temperature range, suggesting that this fast process is related to the strong exciton-trion coherent coupling, in which the energy transition is in the sub-picosecond range\cite{PhysRevLett.112.216804}. The neutral excitons thus act as a valley relaxation channel for trions. The slower valley decay process for trions suggests that there's still residue carrier valley polarization after neutral exciton valley relaxation. The electron (hole) valley relaxation needs spin flips in addition to the intervalley scattering, which is expected to relax in a longer time than that of exciton as the MSS mechanism is absent.

Several mechanisms may be responsible for the electron (hole) valley relaxation, in which spin relaxation is essential, including D'yakonov-Perel' (DP), Elliot-Yafet (EY) and Bir-Aronov-Pikus (BAP) mechanism as well as the hyperfine interaction and magnetic scatterings\cite{Spindynamicsinsemiconductors}. Here the hyperfine interaction can be ruled out since the electron wavefunction hardly overlaps with the nucleus'\cite{ControlofvalleypolarizationinmonolayerMoS2byopticalhelicity}. The magnetic scatters allow an electron to overcome the energy barrier of the large spin splitting in both the valence and conduction bands\cite{ControlofvalleypolarizationinmonolayerMoS2byopticalhelicity}. It has been evidenced that there exist magnetic scatters in TMDs\cite{C2JM15906F,PhysRevB.80.125416,:/content/aip/journal/apl/101/12/10.1063/1.4753797}, but this is very unlikely to be the case for the studied WSe$_2$ flake here, as no obvious Kerr response is observed when pumped by linearly polarized beam (see FIG. \ref{4}a).

The spin relaxation time governed by the DP and EY mechanisms has previously been calculated to be in picosecond range. The spin relaxation time is suggested to decrease with increasing carrier density based on the numerically calculated results\cite{PhysRevB.87.245421,PhysRevB.89.115302,PhysRevLett.111.026601}. The EY mechanism ought to be negligible for the spin relaxation in TMD materials because of the marginal spin mixing\cite{PhysRevB.89.115302}. Previous theoretical studies\cite{zhu2014anomalously,ValleydepolarizationduetointervalleyandintravalleyelectronholeexchangeinteractionsinmonolayerMoS} have suggested that DP mechanism should also be negligible for spin flip along the out-of-plane direction as the mirror symmetry with respect to the plane of W atoms secures a zero out-of-plane crystal electric field. However, as a matter of fact, the mirror symmetry is actually broken since the surface roughness of SiO$_2$ substrates is usually as large as 4-8 \AA \
as previously reported\cite{sahin2013anomalous}, in addition to the possible morphology distortion due to the existence of absorbate or residue glue. Thus it is very likely that there exists a spin-orbit coupling induced in-plane effective magnetic field, which opens an efficient spin relaxation channel as described by the DP mechanism. The slow 
spin relaxation in nanosecond range in monolayer MoS$_2$ reported recently\cite{yang2015long} might be the consequence of the protected mirror symmetry by the better sample crystallization quality grown by chemical vapor deposition.

The BAP mechanism could be enhanced compared to the III-V group semiconductors due to the enhanced electron-hole coupling in 2D limit and the relatively high unintentional doping level\cite{ControlofvalleypolarizationinmonolayerMoS2byopticalhelicity,PhysRevB.87.245421}. So even with larger energy splitting in the valence band, the holes may possess similar valley lifetime with the electrons due to the BAP mechanism, where the spin of a hole may be flipped by the exchange interaction with a conduction electron.

Detailed works such as substrate-engineered and doping controlled trion valley dynamics are needed in the future.

\section{Conclusions}
In conclusion, the valley dynamics of different excitonic states including free excitons and trions in monolayer WSe$_2$ is investigated by the two-color pump-probe TRKR measurement. The valley relaxation times of excitonic states are confirmed to be in picosecond range. The temperature dependent valley relaxation times for both excitons and trions are observed to decrease with increasing temperature, following the dominated MSS mechanism caused by the \emph{e-h} exchange interaction. The trion state is found to be able to preserve the valley polarization more efficiently than the neutral exciton, thus being more promising for valleytronics application.

\section{Acknowledgements}

This work is supported by the National Natural Science Foundation of China (Nos. 11474276, 11274302, 11225421, 11434010 and 11474277). We would like to thank Dr. Hui Zhu on setting up the experimental components.

\nocite{*}
\bibliography{aipsamp}

\end{document}